\documentclass{aastex631}

\usepackage{multirow}
\usepackage{booktabs}
\usepackage{amsmath,amssymb}
\usepackage[T1]{fontenc}
\usepackage{threeparttable}
\usepackage{amsmath}
\usepackage{xcolor}
\usepackage{graphicx}
\usepackage{float}
\usepackage{subfigure}
\usepackage{hyperref}
\usepackage{comment}
\usepackage{url}
\usepackage{bm}
\shorttitle{Shear Particle Acceleration in Structured Gamma-Ray Burst Jets}
\shortauthors{Wang, Huang,\,\& Liang.}

\begin{document}
\title{Shear Particle Acceleration in Structured Gamma-Ray Burst Jets: I. Physical Origin of the Band Function and Application to GRBs 090926A, 131108A, and 160509A}

\correspondingauthor{Xiao-Li Huang, En-Wei Liang}
\email{xiaoli.huang@gznu.edu.cn, lew@gxu.edu.cn}

\author[0009-0001-8025-3205]{Zi-Qi Wang}
\affiliation{Guangxi Key Laboratory for Relativistic Astrophysics, School of Physical Science and Technology, Guangxi University, Nanning 530004, People’s Republic of China}

\author[0000-0002-9725-7114]{Xiao-Li Huang$^{*}$}
\affiliation{School of Physics and Electronic Science, Guizhou Normal University, Guiyang 550025, People’s Republic of China}
\affiliation{Guangxi Key Laboratory for Relativistic Astrophysics, School of Physical Science and Technology, Guangxi University, Nanning 530004, People’s Republic of China}

\author[0000-0002-7044-733X]{En-Wei Liang$^{*}$}
\affiliation{Guangxi Key Laboratory for Relativistic Astrophysics, School of Physical Science and Technology, Guangxi University, Nanning 530004, People’s Republic of China}

\begin{abstract}
The radiation physics of gamma-ray bursts (GRBs) remains an open question. Based on the simulation analysis and recent observations, it was proposed that GRB jets are composed of a narrow ultra-relativistic core surrounded by a wide sub-relativistic cocoon. We show that emission from the synchrotron radiations and the synchrotron self-Compton (SSC) process of shear-accelerated electrons in the mixed jet-cocoon (MJC) region and internal-shock-accelerated electrons in the jet core is potentially explained the spectral characteristics of the prompt gamma-rays. Assuming an exponential-decay velocity profile, the shear flow in the MJC region can accelerate electrons up to $\gamma_{\rm e,\max} \sim 10^4$ for injected electrons with $\gamma_{\rm e,inject}=3 \times 10^2$, if its magnetic field strength ($B_{\rm cn}$) is $100$ G and its inner-edge velocity ($\beta_{\rm cn, 0}$) is 0.9c. The cooling of these electrons is dominated by the SSC process, and the emission flux peaks at the keV band. In addition, the energy flux of synchrotron radiations of internal- shock-accelerated electrons ($\gamma_e=10^{4}\sim 10^{5}$) peaks at around the keV$-$MeV band, assuming a bulk Lorentz factor of 300, a magnetic field strength of $\sim 10^{6}$ G for the jet core. Adding the flux from both the jet core and the MJC region, the total spectral energy distribution (SED) illustrates similar characteristics as the broadband observations of GRBs. The bimodal and Band-Cut spectra observed in GRBs 090926A, 131108A, and 160509A can be well fit with our model. The derived $B_{\rm cn}$ varies from 54 G to 450 G and $\beta_{\rm cn,0}=0. 83\sim 0.91$c.   
\end{abstract}

\keywords{Gamma-ray bursts (629); Non-thermal radiation sources (1119)}

\section{Introduction} \label{sec:intro}
Gamma-ray bursts (GRBs) are extreme electromagnetic events in the universe. Extensive observations with gamma-ray missions have accumulated a large sample of GRB spectra in the keV$-$MeV$-$GeV bands. Specifically, the GRB spectrum in the keV$-$MeV band observed with the Burst And Transient Source Experiment (BATSE, 20$-$1000 keV) on-board the Compton Gamma-Ray Observatory (CGRO) mission is typically fitted with a smooth broken power-law function, known as the Band function \citep{1993ApJ...413..281B,2006ApJS..166..298K}.
This characteristic is also confirmed with observations of the Gamma-Ray Burst Monitor (GBM,15$-$1000 keV) on-board the {\em Fermi} Gamma-ray Space Telescope \citep{2011ApJ...730..141Z,2021ApJ...913...60P}.
Furthermore, the joint spectra of some GRBs observed with both the GBM and the Large Area Telescope (LAT) on-borad the {\em Fermi} in the 8 keV$-$300 GeV range exhibit either a bimodal structure (such as GRB 090926A;\citealp{2011ApJ...729..114A,2017A&A...606A..93Y}) or a Band function with cut-off (the so-called Band-Cut function; \citealp{2013ApJS..209...11A,2017ApJ...844...56T}). It is indicated that the spectra of generic GRBs in the keV$-$GeV band maybe embed an extra component beneath the Band function \citep{2011ApJ...730..141Z}. 

It is suggested that GRBs originate from ultra-relativistic jets powered by collapses of massive stars or mergers of compact object \citep{1992ApJ...395L..83N,1993ApJ...405..273W,2002ARA&A..40..137M,2004RvMP...76.1143P,2015PhR...561....1K}. Within the framework of the standard jet model, a photosphere stage and an internal dissipation stage (internal shocks, internal magnetic processes, etc.) are expected \citep{1986ApJ...308L..43P,1994ApJ...430L..93R,1998MNRAS.296..275D,2000ApJ...530..292M,2003astro.ph.12347L,2011ApJ...726...90Z}. Consequently, the predicted prompt emission spectrum may consist of a thermal component from the photosphere emission and a non-thermal component from the synchrotron (Syn) and/or the Inverse Compton (IC) emission of the accelerated electrons \citep{1986ApJ...308L..47G,1994ApJ...430L..93R,1998ApJ...494L.167P}. Furthermore, the propagation of a relativistic jet through the surrounding medium drives a bow shock, forming a cocoon of shocked material that deposits significant energy \citep{2000ApJ...531L.119A,2003ApJ...586..356Z,2011ApJ...740..100B,2019MNRAS.490.4271M}. When the jet breakout, the cocoon material erupts and disperses radially and axially, resulting in the formation of a jet-cocoon structure \citep{2002MNRAS.337.1349R,2007ApJ...665..569M}. Such a jet-cocoon structure has been the focus of intensive investigations \citep{2005ApJ...629..903L,2006ApJ...652..482P,2017ApJ...834...28N,2019ApJ...881...89L}. 
GRBs 170817A and 221009A are two representative cases for revealing the GRB ejecta structure. The observed short GRB 170817A, associated with the binary neutron star merger gravitational wave (GW) signal GW170817 \citep{2017PhRvL.119p1101A,2017ApJ...848L..13A}, has been suggested to be attributable to off-axis observations of a structured ejecta with a large viewing angle \citep{2018MNRAS.476.1191B,2018Natur.561..355M,2018MNRAS.479..588G,2019ApJ...871..123F}.
On the other hand, the broadband afterglow lightcurves of long GRB 221009A are fitted by considering a core-wing configuration \citep{2023SciA....9I1405O,2023MNRAS.522L..56S,2023MNRAS.524L..78G,2024ApJ...962..115R}.

Relativistic hydrodynamic and magnetohydrodynamic simulations reveal a radial velocity distribution within the jet-cocoon outflow, that is known as shear flow \citep{2000ApJ...531L.119A,2008MNRAS.388..551T,2013ApJ...777..162M,2021ApJ...915L...4G}. Particles could be accelerated within the shear flow\citep{1981SvAL....7..352B,1989ApJ...340.1112W,2004ApJ...617..155R,2018ApJ...855...31W}. The shear-accelerated electrons may be expected as a potential contributor to the prompt emission of GRBs. As a result, the prompt gamma-ray spectrum may be shaped by two distinct electron populations: one accelerated by internal shocks via the Fermi acceleration mechanism within the jet, and the other by the shear acceleration mechanism within the mixed jet-cocoon (MJC) region. In this paper, we investigate a comparative analysis of the GRB radiation in the framework of synchrotron and SSC emissions of electrons accelerated in the jet core and the MJC region. We present our model in Sec.~\ref{sec:Shear acceleration} and apply this model to fit the spectra of GRBs 090926A, 131108A, and 160509A in Sec.~\ref{sec:Case Study}. The spectra of these GRBs display a bimodal feature or a deviation from the standard Band function profile. The summary and discussion are presented in Sec.~\ref{sec:Summary}. Throughout this paper, we employ a Hubble constant of $H_0=71\ \mathrm{km} \mathrm{s}^{-1}\,\,\mathrm{Mpc}^{-1}$, and the cosmological parameters of $\varOmega _M=0.27$ and $\varOmega _{\Lambda}=0.73$.

\section{Model} \label{sec:Shear acceleration}

\subsection{Jet-Cocoon Structure} \label{subsec:jet-cocoon Structure}
\begin{figure}[htbp!]
    \centering
    \includegraphics[width=0.45\textwidth]{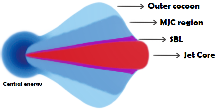}
    \includegraphics[width=0.2\textwidth]{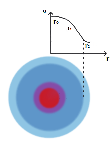}
    \caption{The schematic diagram of the Jet-Cocoon structure.}
    \label{fig:Jet-Cocoon structure}
\end{figure}

Motivated by the results of numerical simulations and theoretical calculations, we conceptualize the GRB ejecta as a jet-cocoon structure, as illustrated in Figure~\ref{fig:Jet-Cocoon structure}. This structure consists of three distinct regions: an ultra-relativistic narrow jet core region with an uniform velocity profile ($r<r_0$), a sub-relativistic mixed jet-cocoon region with decreasing velocity as a function of radial radius ($r_0<r<r_2$), and an outer cocoon region with an uniform velocity profile ($r>r_2$). Moreover, particle-in-cell (PIC) simulations have demonstrated significant particle acceleration at the shear boundary layer (SBL) \citep{2014NJPh...16c5007A}. We also illustrate the SBL as a thin layer at $r_0\lesssim r\lesssim r_1$. Hereinafter, variables with the subscripts ``jet'' and ``cn'' refer to the jet and cocoon regions, respectively.

We set the distance of the emitting regions of both the jet and MJC region from the central engine as $R$ \citep{2002MNRAS.337.1349R,2011ApJ...726...90Z,2015AdAst2015E..22P}. 
In the framework of jet-cocoon structure, we postulate that the GRB ejecta during the prompt emission phase remains in a steady-state scenario, with no significant density variation and lateral (radial) expansion. We consider that the velocity $\boldsymbol{u}$ of MJC region is along the direction of the jet axis (i.e. $\boldsymbol{u}=u_{\rm cn}(r) \boldsymbol{e}_{z}$), and the velocity profile is modeled as an exponential-decay function 

\begin{equation}
u_{\rm cn}(r)=\beta _{\rm cn,0}e^{-k}, \ \  k = \frac{r \ln (\beta_{\rm cn,0} / \beta_{\rm cn,2})}{r_{2}}
\label{eq:u(r)}, 
\end{equation}
where $r$ is the radial distance from the jet axis, $u_{\mathrm{cn}}$ is the outflow velocity in units of the light speed c, $\beta_{\rm cn,0}$ and $\beta_{\rm cn,2}$ are the fluid velocities at $r_0$ and $r_2$, respectively. Theoretically, the velocity of the outer cocoon ejecta should not exceed the local sound speed. In this work, we constrain the boundary velocity to $\beta_{\rm cn,2} < 1/\sqrt{3}$ \citep{2002MNRAS.337.1349R,2010ApJ...709L..83M}. We assume an uniform magnetic field strength within the jet core and the MJC region.

\subsection{Particle Acceleration in the MJC region} \label{subsec:Shear acceleration model}
Particles could be accelerated through the shear acceleration mechanism in the MJC region. This process involves the coupling of the energetic particles and the shear force in the outflow due to cosmic-ray viscosity, as well as the scattering process arising from magnetic field irregularities embedded in the background outflow \citep{1981SvAL....7..352B,1989ApJ...340.1112W,2002A&A...396..833R,2004ApJ...617..155R,2018ApJ...855...31W}. 

As mentioned above, PIC simulations show significant particle acceleration at the SBL, attributed to the combined effects of instabilities and electromagnetic fields \citep{2012ApJ...746L..14A,2017ApJ...847...90L}. This also leads to notable particle accumulation within the SBL. The SBL sustains a prolonged particle acceleration process and efficiently energizes the majority of particles. For leptons, this mechanism maybe achieve energies characterized by $\gamma_{\mathrm{eff}}\sim \Gamma_{\rm jet}$, proceeding in a strongly an-isotropic manner \citep{2017ApJ...847...90L}. Therefore, we designate the SBL as an electron injection layer of the MJC region and assume the injected electrons are a mono-energetic population with $\gamma_{e, \mathrm{inject}} = \gamma_{\rm reff}$. 

The injected electrons are further accelerated in the MJC region. We focus on the strong scattering limit case, where collisions are sufficiently effective to restore isotropy \citep{2005ApJ...632L..21R}. 
Based on the strong particle scattering assumption, \cite{2018ApJ...855...31W} provides a steady-state solution for particle acceleration in the relativistic shear flow within the isotropic diffusion model. When the scattering wave frame is taken to coincide with the comoving fluid frame, the transport equation for the isotropic shear-accelerated particles distribution function $f_0\left( x^{\alpha},p \right) $ $(x^{\alpha}=(ct,x,y,z))$ in momentum space can be expressed as \citep{1975MNRAS.172..557S,1989ApJ...340.1112W,2018ApJ...855...31W}
\begin{equation}
\begin{split}
    &\nabla _{\mathrm{\alpha}}\left[ cu^{\alpha}f_0-\kappa \left( \eta ^{\alpha \beta}+u^{\alpha}u^{\beta} \right) \left( \frac{\partial f_0}{\partial x^{\beta}}-\dot{u}_{\beta}\frac{\left( p^0 \right) ^2}{p}\frac{\partial f_0}{\partial p} \right) \right] \\
    &+\frac{1}{p^2}\frac{\partial}{\partial p}\left[ -\frac{p^3}{3}{cu_{;\beta}^{\beta}}f_0+p^3\left( \frac{p^0}{p} \right) ^2 \right. \\
    &\left. \times \kappa \dot{u}^{\beta}\left( \frac{\partial f_0}{\partial x^{\beta}}-\dot{u}_{\beta}\frac{\left( p^0 \right) ^2}{p}\frac{\partial f_0}{\partial p} \right) -\Lambda  \tau p^4\frac{\partial f^0}{\partial p} \right] =Q,
    \label{eq:transport equation}
\end{split}
\end{equation}
where $\eta ^ {\alpha\beta}$ is the Minkowski metric (scripts $\alpha,\beta=0,1,2,3$), $u^{\alpha}$ denotes the fluid velocity four-vector, and $\dot{u}$ represents the acceleration vector of the fluid, $u_{;\beta}^{\beta}$ represents the covariant derivative. $p$ is the comoving particle momentum, $p^0=E/c$ is the zeroth component of the particle momentum four-vector in the fluid frame. $\tau$ is the scattering or collision timescale. $\kappa$ is the particle diffusion coefficient ( $\kappa ={{{v}^2\tau}/{3}}$, where $v$ is the particle speed in the comoving frame). $Q$ represents particle source, and $\Lambda$ quantifies the viscous energization coefficient. Under the strong scattering limit, $\Lambda$ is given by
\begin{equation}
\Lambda  =\frac{c^2}{30}\sigma _{\alpha \beta}\sigma ^{\alpha \beta},
\label{eq:3}
\end{equation}
where $\sigma _{\alpha \beta}$ is the shear tensor of the background flow \citep{1989ApJ...340.1112W,2018ApJ...855...31W} .

We consider particles to be bounded within the MJC region ($r_0 < r < r_2$) where the outflow is treated as steady-state and incompressible. Particles are injected into the region at $r = r_1\ (r_1 \gtrsim r_0)$ with the momentum $p_0 \sim \gamma_{e, \mathrm{inject}} / m_e c$, and escape from the acceleration region at $r = r_2$.
The particle source term $Q$ is given by
\begin{equation}
Q=\frac{1}{2\pi r_1}\frac{N_0}{4\pi{p_0}^2}\delta \left( p-p_0 \right) \delta \left( r-r_1 \right),
\label{eq:4}
\end{equation}
where $N_0$ is the initial distribution of injected electrons. In accordance with the strong scattering limit, we assume a relatively weak average magnetic field and strong turbulence within the MJC region \citep{2001A&A...369..694S,2005ApJ...632L..21R,2012ApJ...744...32Z}.

Then the covariant derivative $u_{;\beta}^{\beta}$ can be specified as $u_{;\beta}^{\beta} = 0$, and the viscous energization coefficient $\Lambda$ has the form
\begin{equation}
\Lambda =\frac{\mathrm{c}^2}{15}\Gamma _{\mathrm{cn}}^{4}\left( \frac{du_{\mathrm{cn}}}{dr} \right) ^2.
\label{eq:Lambda}
\end{equation} 
Accordingly, in the steady-state relativistic MJC region within the GRB environment, the transport equation (Eq.~\ref{eq:transport equation}) for the distribution function of shear-accelerated electrons can be recast as
\begin{equation}
-\frac{1}{r}\frac{\partial}{\partial r}\left( \kappa r\frac{\partial f_0}{\partial r} \right) -\frac{c^2}{15}\frac{\Gamma _{\mathrm{cn}}^{4}}{p^2}\left( \frac{du_{\mathrm{cn}}}{dr} \right) ^2\frac{\partial}{\partial p}\left( p^4\tau \frac{\partial f_0}{\partial p} \right) = \frac{N_0 \delta \left( p-p_0 \right) \delta \left( r-r_1 \right)}{8\pi^2{p_0}^2 r_1} .
\label{eq:simplify1}
\end{equation}
The analytical solution of Eq.~\ref{eq:simplify1} for the shear-accelerated electron distribution function $f_0$ can be formulated as \citep{2018ApJ...855...31W}  
\begin{equation}
\begin{split}
    f_0=&\frac{15}{8 \pi^{2} \left( \xi _0-\xi _2 \right)\left| \frac{d\xi}{dr}|_{r_1} \right|r_1}\left( \frac{N_0}{{p_0}^{3}c^2\tau _0} \right) \exp \left[ -\frac{\left( 3+\alpha \right) T}{2} \right] \\
    &\times \sum_{n=0}^{\infty}{\frac{1}{y_n}\sin \left[ \left( n+\frac{1}{2} \right) \pi w_1 \right]}\sin \left[ \left( n+\frac{1}{2} \right) \pi w \right] \exp \left( -y_n\left| T \right| \right) ,
\label{eq:f0}
\end{split}
\end{equation}
in which  
\begin{equation}
\xi \left( r \right) =\frac{1}{2}\ln \left( \frac{1+u_{\mathrm{cn}}}{1-u_{\mathrm{cn}}} \right) , \ \  w \equiv \frac{\xi -\xi _2}{\xi _0-\xi _2}, \ \ T=\ln \left( \frac{p}{p_0} \right) 
\label{eq:6}
\end{equation}
\begin{equation}
y_n=\left[ \frac{5\pi ^2\left( 2n+1 \right) ^2}{4\left( \xi _0-\xi _2 \right) ^2}+\frac{\left( 3+\alpha \right) ^2}{4} \right] ^{{{1}/{2}}}, \ \  n=0,1,2,\dots ,
\label{eq:7}
\end{equation}
where the subscripts 0, 1, and 2 represent the physical quantities at $r=r_0$, $r=r_1$, and $r=r_2$, respectively. $\tau_0$ is the initial scattering timescale and $\alpha = 2 - q$ is a constant that determines the momentum dependence of the mean scattering time $\tau$. $q$ is the spectral index of the turbulence model. Employing the Kolmogorov turbulence model, we ascertain the wave number spectral index as $q=5/3$, corresponding to $\alpha=1/3$ \citep{1941DoSSR..30..301K}. 

In the frame of the quasi-linear theory, the scattering time ($\tau$) can be consistent with the mean free path formula ($\lambda$) \citep{2017ApJ...842...39L,2018ApJ...855...31W}. $\tau$ and $\lambda$ are given by
\begin{equation}
\tau \left( r,p \right) =\tau _0\left( \frac{p}{p_0} \right) ^{\alpha}\frac{r_1\xi^{'}\left( r_1 \right)}{r\xi^{'}\left( r \right)},\ \ \lambda=\frac{r_{g}^{2-q}\ell_{b}^{q-1}}{c\chi}N\left( q \right), 
\label{eq:8}
\end{equation}
where $N\left( q \right) ={{3}/{\left[ \left( 2-q \right) \left( 4-q \right) \right]}}$, $r_g=pc/\left( e B_{\rm cn} \right)$ is the gyroradius of particle, $\ell_b = 1/k_b$ corresponds roughly to the correlation length of the turbulence, and
\begin{equation}
\chi =\left( \frac{\delta B_{\rm cn}}{B_{\rm cn}} \right) ^2\frac{1}{\Phi \left( k_b,k_d \right)},\ \ \Phi \left( k_b,k_d \right) =
	1+\frac{1-\left( k_b/k_d \right) ^{1-q}}{q-1}, 
\end{equation}
in which $\delta B_{\rm cn}$ is the magnetic field fluctuation perpendicular to the magnetic field $B_{\rm cn}$ ($\delta B_{\rm cn} \ll B_{\rm cn}$; \citealp{2004JGRA..109.4107Z,2017ApJ...842...39L}), $k_b$ and $k_d$ are the range of resonant wave number for interactions. In our analysis, the parameters $k_b$ and $k_d$ ideally keep as $k_b=10^{-13}$ cm and $k_d=10^{-2}$ cm, respectively \citep{2003matu.book.....B,2018ApJ...855...31W}. With $\tau=\lambda$, $\tau_0$ can be inferred as  
\begin{equation}
\tau _0=\left( \frac{\ell _b}{c} \right) \frac{N\left( 2-q \right)}{\chi \left( r_1 \right)}\left( \frac{p_0c}{\ell _b e B_{\rm cn}} \right) ^{\alpha}.
\end{equation}

\begin{figure}[htbp!]
    \centering
    \includegraphics[width=0.35\textwidth]{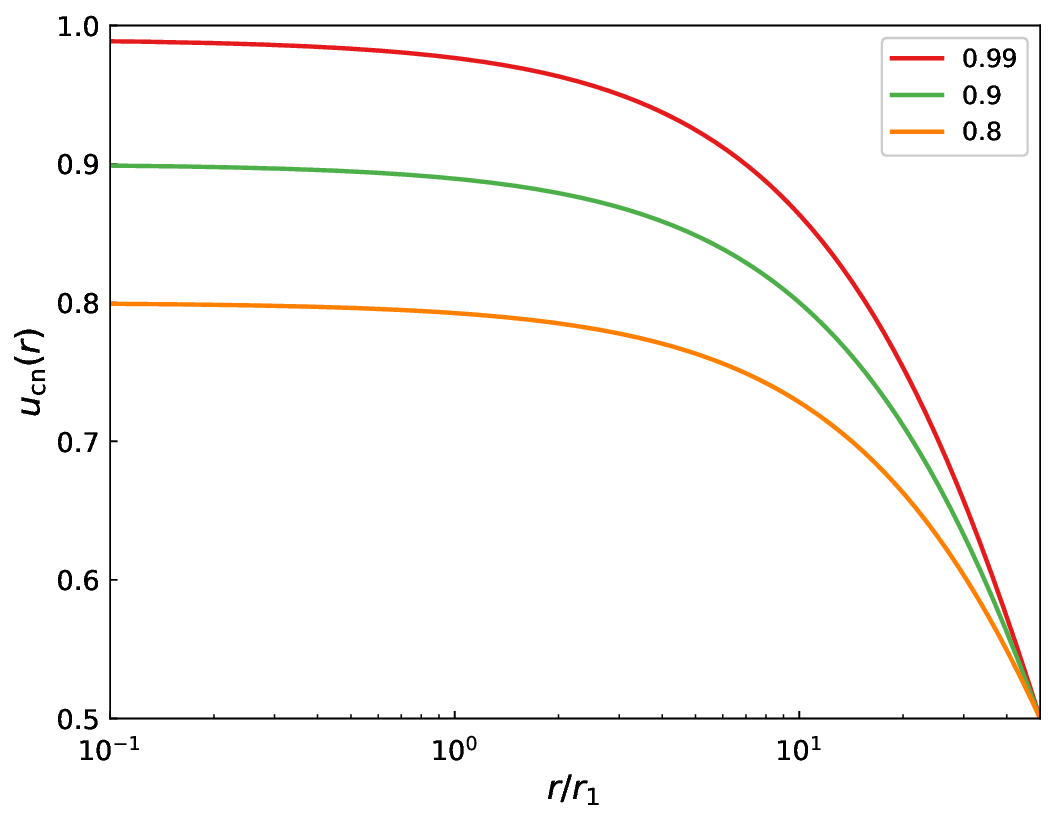}
    \includegraphics[width=0.365\textwidth]{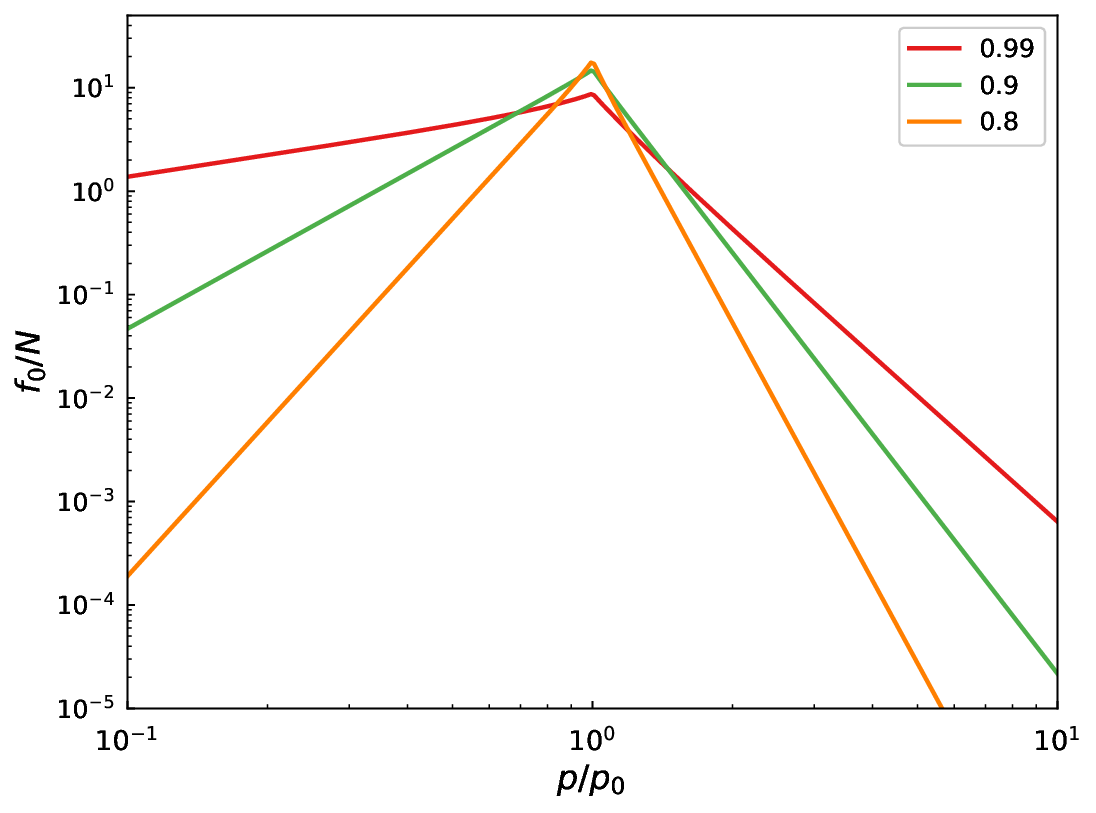}
    \caption{{\rm Left panel--} Velocity profiles of the MJC region as an exponential-decay function of radius, with initial velocities of $\beta_{\rm cn,0}=0.99, 0.9, 0.8$. {\rm Right panel--}  Distributions of shear-accelerated electrons for exponential-decay velocity profiles as shown in the left panel, where $N=N_0/\left( {p_0}^3c^2\tau _0r_1\left| \frac{d\xi}{dr}|_{r_1} \right| \right)$.}
    \label{fig:distribution}
\end{figure}

We calculate shear-accelerated particle distributions within the MJC region. The parameters of the MJC region are set as follows: the distance to the central engine $R=10^{15}$ cm, the full opening angles $\theta_{\mathrm{cn}} = 0.7$ rad, the magnetic field strength $B_{\mathrm{cn}}=100$ G \citep{2001A&A...369..694S,2006ApJ...652..482P}, and $\gamma_{e, \mathrm{inject}}=300$. Figure~\ref{fig:distribution} illustrates the velocity profiles for various $\beta_{\rm cn,0}$ values and the corresponding electron distributions accelerated via the shear acceleration mechanism. One can find that the particle distribution is broader for larger $\beta_{\rm cn, 0}$. At the high momentum band $(p>p_0)$, the shear-accelerated particle distribution exhibits a power-law decay behavior, which is characterized by $f_0\propto p^{-\mu _{\infty}}$. Based on Eq.~(\ref{eq:f0}), the spectral index $\mu _{\infty}$ is given by \citep{2018ApJ...855...31W}
\begin{equation}
\mu _{\infty}=\left( \frac{5\pi^{2}}{4\left( \xi _0-\xi _2 \right)^{2}}+\frac{\left( 3+\alpha \right) ^2}{4} \right) ^{1/2}+\frac{\left( 3+\alpha \right)}{2} .
\label{eq:10}
\end{equation}  
At the low energy band ($p<p_0$), the spectrum initially presents as an inverse power-law behavior, described by $f_0\propto p^{-\mu _0}$ when $p\rightarrow 0$. The power law index $\mu_0$ is defined as
\begin{equation}
\mu _0=\left( \frac{5\pi^{2}}{4\left( \xi _0-\xi _2 \right)^{2}}+\frac{\left( 3+\alpha \right) ^2}{4} \right) ^{1/2}-\frac{\left( 3+\alpha \right)}{2}=\mu _{\infty}-\left( 3+\alpha \right) .
\label{eq:11}
\end{equation} 
The particle distribution resulting from shear acceleration is dramatically different from the typical first-order Fermi acceleration ($f_{\mathrm{st}}\propto p^{-q_{\mathrm{st}}}$) and second-order Fermi acceleration ($f_{\mathrm{nd}}\propto p^{-2}e^{3-q_{\mathrm{nd}}}$) \citep{1990ApJ...360..702E,2005PhRvL..94k1102K,2012ApJ...746..164M}.

The maximum shear-accelerated electron Lorentz factor $\gamma_{M, \rm cn}$ is restricted by $t_{\mathrm{acc,cn}} = t_{\mathrm{rad,cn}}$, where $t_{\mathrm{acc,cn}}$ is the shear acceleration timescale and $t_{\mathrm{rad,cn}}$ is the cooling timescale via the synchrotron radiation and the synchrotron self-Compton ($\rm SSC$) process. The $t_{\mathrm{acc,cn}}$ is estimated as  
\citep{2018ApJ...855...31W}
\begin{equation}
t_{\mathrm{acc},\mathrm{cn}}=\frac{p}{\left< \Delta p/\Delta t \right>}=\frac{15}{\left( 4+\alpha \right) {\Gamma _{\mathrm{cn}}}^4\left( du_{\rm cn}(r)/dr \right) ^2\tau} ,
\end{equation}
where $\Gamma_{\rm cn}$ is the Lorentz factor of the cocoon region, and $\tau$ is given by Eq.~\ref{eq:8}. The $t_{\mathrm{rad,cn}}$ value is calculated with \citep{2009ApJ...703..675N}
\begin{equation}
t_{\mathrm{rad,cn}}=t_{\mathrm{Syn_{cn}}}+t_{\rm {SSC_{cn}}}=\frac{6\pi m_ec}{\gamma _{e,\rm cn}\sigma _{\mathrm{T}}B_{\mathrm{cn}}^2\left( 1+\mathrm{Y_{\rm cn}} \right)} ,
\end{equation}
where $Y_{\rm cn}$ is the Compton parameter, defined as the ratio of the $\rm SSC_{cn}$ power to the $\rm Syn_{cn}$ power. The radial scale of the MJC region is estimated as $r_{\rm cn}\simeq \theta _{\rm cn}/ 2 \times R=3.5 \times 10^{14}$ cm. Taking $\beta_{\rm cn,0} = 0.9$, $B_{\rm cn} = 100$ G, $\gamma_{e, \mathrm{inject}} = 3 \times 10^{2}$, we have $\gamma_{\rm M, cn} \sim 10^{4}$. These electrons should also be confined in the acceleration region. This requires that the gyroradius ($r_g$) of the electron is smaller than the maximum wavelength ($\ell_b$) for particle scattering. Considering that particle scattering is primarily influenced by turbulence within the inertial range, we have $\ell_b =\eta r_{\rm cn}$, where $\eta \lesssim 1$ \citep{2017ApJ...842...39L,2018ApJ...855...31W}. Based on above parameters, the $r_g$ of electrons with $\gamma_{e, \rm cn}=10^4$ is $\sim 10^{13}$ cm. Therefore, we have $\ell_b>r_g$. In addition, we also estimate the dynamic timescale of the injected electrons as
$t_{\rm dyn}\sim {r_{\rm cn}}/{c\beta _{\rm cn,0}}\sim 10^4$ s. It is evident that $t_{\mathrm{dyn}}$ is much larger than $t_{\mathrm{acc,cn}}$. These results suggest that the shear acceleration process operates efficiently within the MJC region.

\subsection{Radiation Mechanism and Spectral Energy Distribution}
\begin{figure}[htpb!]
    \centering
    \subfigbottomskip=-3pt
    \subfigcapskip=-5pt
    \subfigure[]{
        \label{fig:spectrum-a}
        \includegraphics[width=0.35\textwidth]{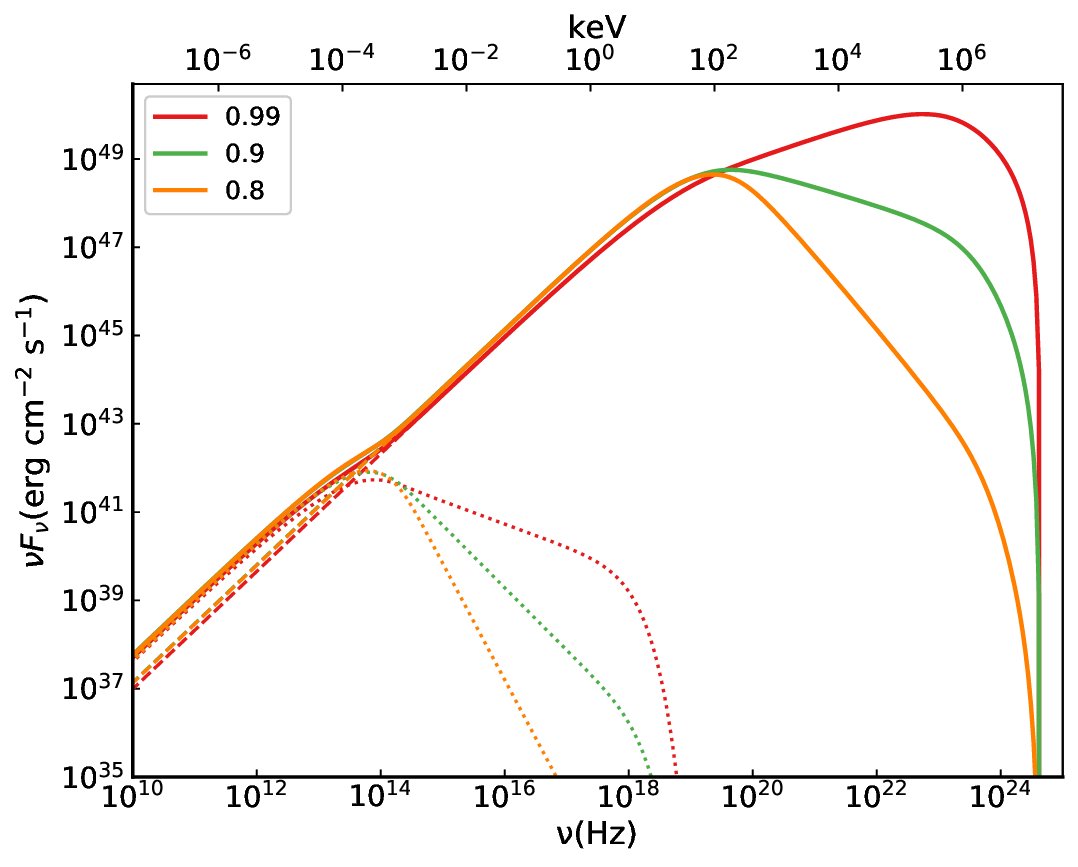}}
    \quad
    \subfigure[]{
        \label{fig:spectrum-b}
        \includegraphics[width=0.35\textwidth]{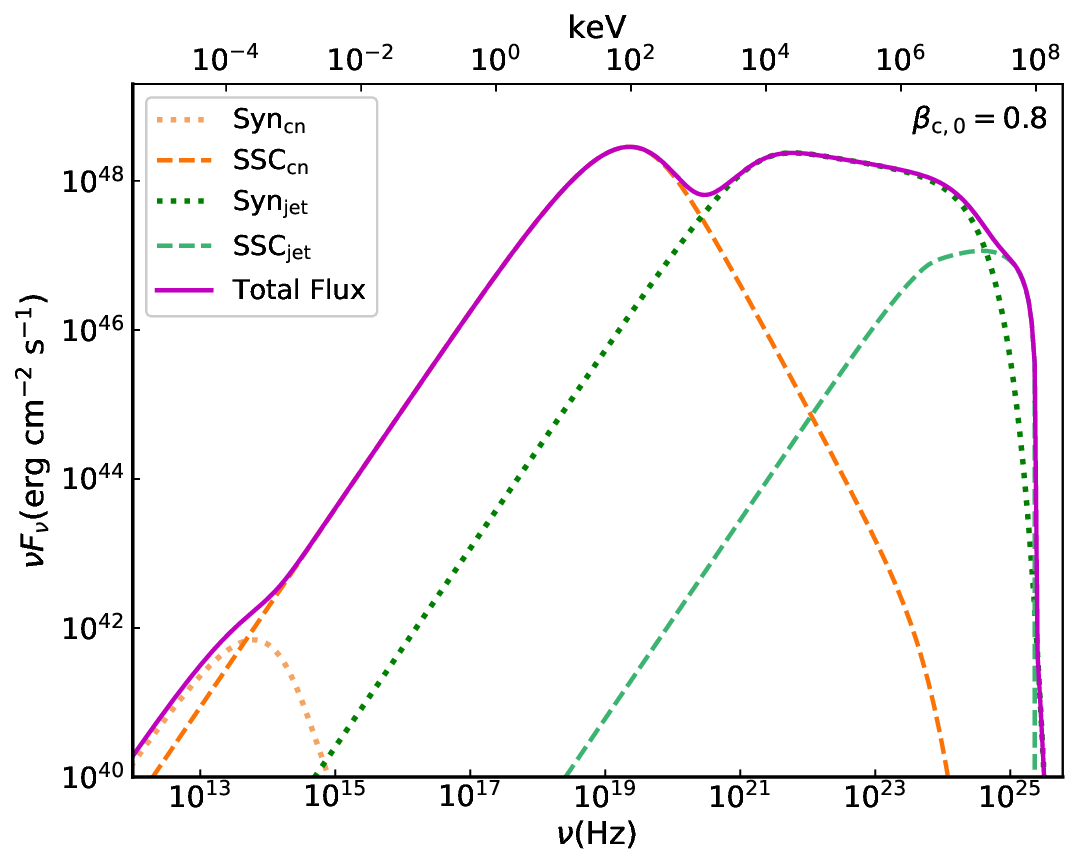}}

    \subfigure[]{
        \label{fig:spectrum-c}
        \includegraphics[width=0.35\textwidth]{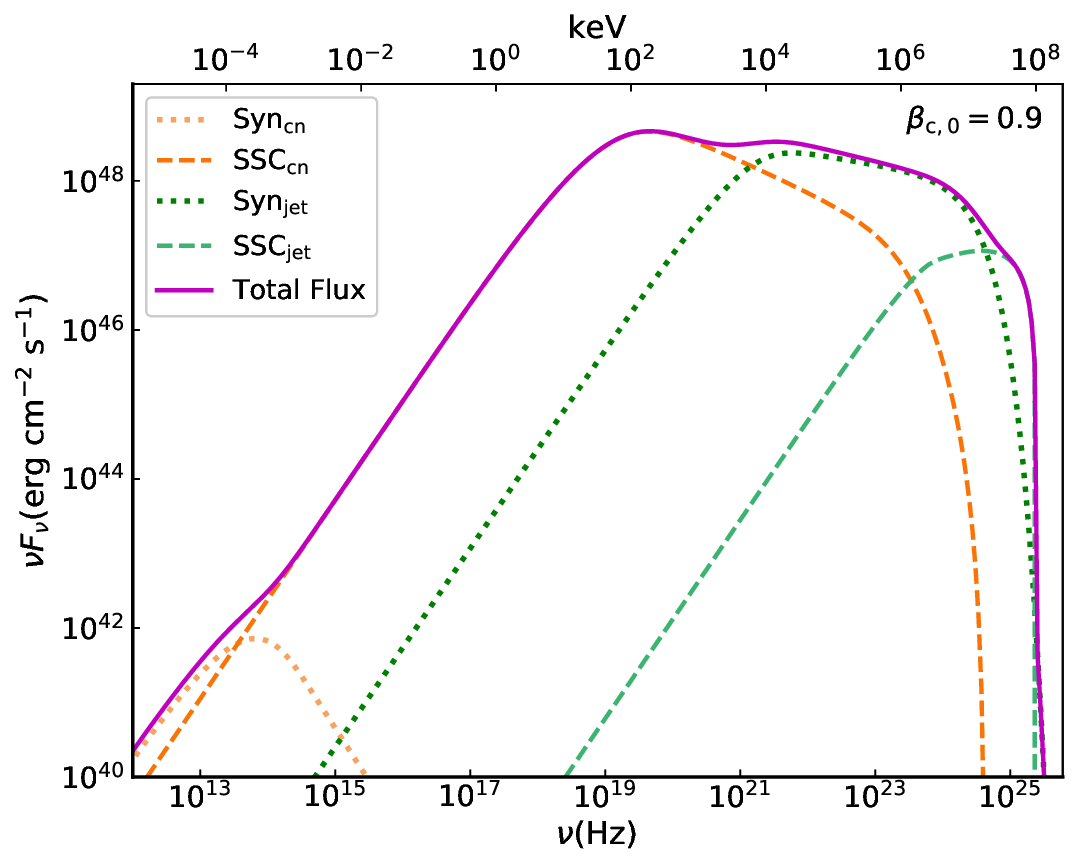}}
    \quad
    \subfigure[]{
        \label{fig:spectrum-d}
        \includegraphics[width=0.35\textwidth]{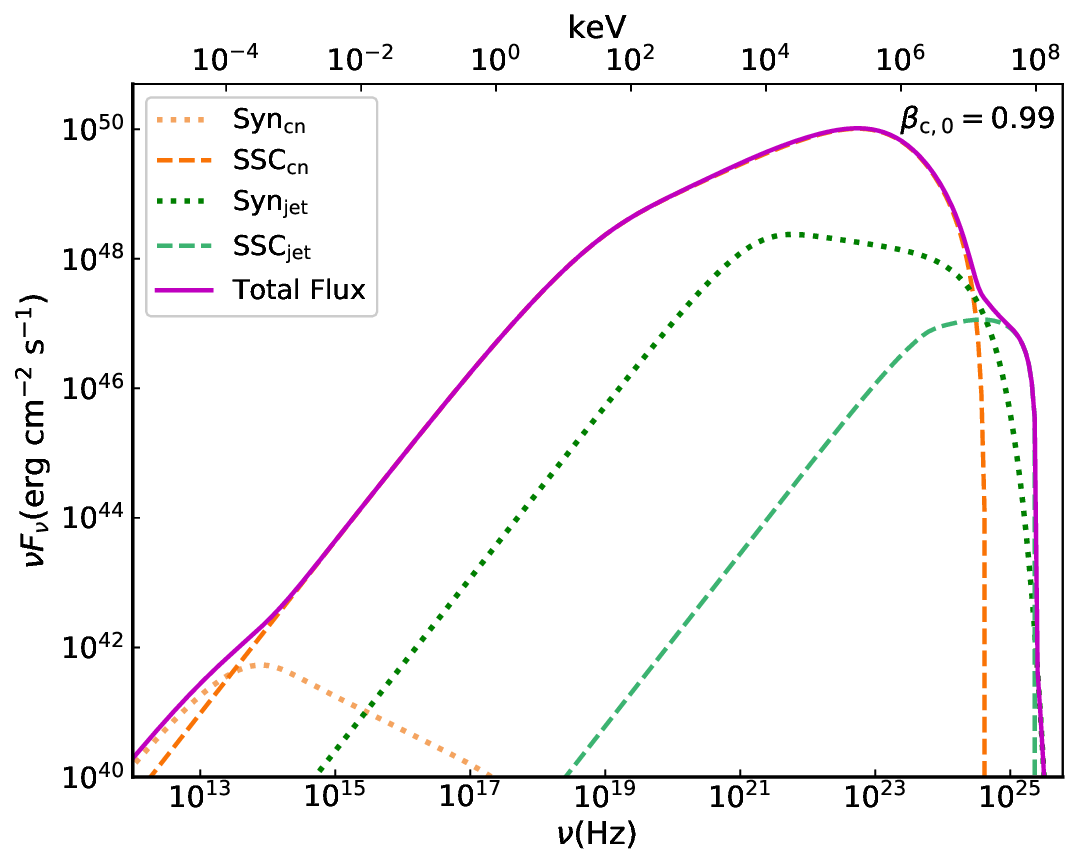}}
    \caption{Panel (a)--The SEDs of the shear-accelerated electrons, with the electron distribution corresponding to Figure~\ref{fig:distribution}. Panels (b, c, d)-- The synthetic SEDs of emissions from the MJC region and the jet core.}
    \label{fig:spectrum}
\end{figure}

The electrons accelerated via the shear acceleration mechanism within the MJC region and through internal shocks in the jet core region are cooled by both the $\rm Syn$ radiation and the $\rm SSC$ process. The parameters of the MJC region and the corresponding shear-accelerated electron distributions are taken as discussed above. We assume the parameters of the jet core as follows: the distance to the central engine $R=10^{15}$ cm, bulk Lorentz factor $\Gamma_{\rm jet}=300$, full opening angle $\theta_{\rm jet}=0.07$ rad, and the magnetic field strength $B_{\rm jet}=10^{6}$ G \citep{2017ApJ...837...33B}. The shock-accelerated electron distribution as a broken power-low function of the electron Lorentz factor in the jet ($\gamma_{e, \rm jet}$ ) is taken as 
\begin{equation}
\frac{dN_{e,\mathrm{jet}}}{d\gamma _{e,\mathrm{jet}}}\propto \begin{cases}
	\gamma _{e,\mathrm{jet}}^{-2}&		\gamma _{m,\mathrm{jet}}\leqslant \gamma _{e,\mathrm{jet}}\leqslant \gamma _{\rm b,jet}\\
	\gamma _{e,\mathrm{jet}}^{-p_{\mathrm{jet}}-1}&		\gamma _{\rm b,jet}<\gamma _{e,\mathrm{jet}}\leqslant \gamma _{\rm M,jet}\\
\end{cases},
\end{equation}
where $p_{\rm jet}$ is the spectral index of electrons accelerated through internal shocks, and $\gamma _{m, \rm jet}$, $\gamma _{\rm b, jet}$, and $\gamma _{\rm M, jet}$ are the minimum Lorentz factor, the break Lorentz factor, and the maximum Lorentz factor of the electrons, respectively. We set $p_{\rm jet} = 2.3$, $\gamma _{m, \rm jet} = 5 \times 10^{3}$, $\gamma _{\rm b, jet} = 1 \times 10^{4}$, and $\gamma _{\rm M, jet} = 2 \times 10^{5}$. Assuming a zero viewing angle to the jet core axis, we calculate the spectral energy distributions (SEDs) of the emission from the MJC region. The results are shown in Figure~\ref{fig:spectrum-a}. One can find that the SSC emission component ($\rm SSC_{cn}$-component) dominates the whole SEDs. More interestingly, the Syn radiation component ($\rm Syn_{cn}$-component) makes a comparable contribution at the low-frequency end ($\nu<10^{14}$ Hz). The gamma-ray flux at $\nu>10^{19}$ Hz is sensitive to $\beta_{\rm cn,0}$.  

The panels b, c, and d of Figure~\ref{fig:spectrum} illustrate the radiation SEDs of the jet core and the MJC region with different $\beta_{\rm cn,0}$ values. For a sub-relativistic MJC region ($\beta_{\rm cn,0}=0.8$ and $\beta_{\rm cn,0}=0.9$), the SED shape as a simi-Band function in the keV$-$MeV band, as usually observed with BTASE \citep{2000ApJS..126...19P}. Due to the contribution of the $\rm SSC_{ cn}$-component, the spectral index in the low-energy band is harder than the prediction of the synchrotron emission model. One can find that the predicted SED in the keV$-$MeV$-$GeV band exhibits either a Band-cut function or a saddle shape. Similar SEDs are indeed found in some GRBs observed with the GBM and LAT. In addition, an X-ray excess over the Band-Cut function around $10^{18-19}$ Hz (several to tens of keV) is found, analogous to the excess observed with the BATSE in the 7$-$20 keV band \citep{2000ApJS..126...19P} and with the Swift in the 2$-$10 keV band \citep{2014ApJ...795..155P}. This excess is attributed to the peak of the $\rm SSC_{cn}$-component. Thirdly, the $\rm Syn_{cn}$-component contributes an Inferred-optical flash that may be called as the prompt IR-optical emission. The peak frequency of this flash depends on the magnetic field strength in the MJC region and the initially injected energy of electrons. For the case of a middle-relativistic MJC region ($\beta_{\rm cn,0}=0.99$) as shown in the panel(d) of Figure~\ref{fig:spectrum}, the emission originating from the MJC region may dominate the overall SED across a broad range, extending from the optical to the sub-TeV energy bands. The SED in the keV$-$GeV band still shows up as the shape of the Band function. The sub-TeV emission is attributed to the $\rm SSC_{jet}$-component.  

\section{Case Study} \label{sec:Case Study}
\begin{figure}[htbp!]
    \centering
        \includegraphics[width=0.55\linewidth]{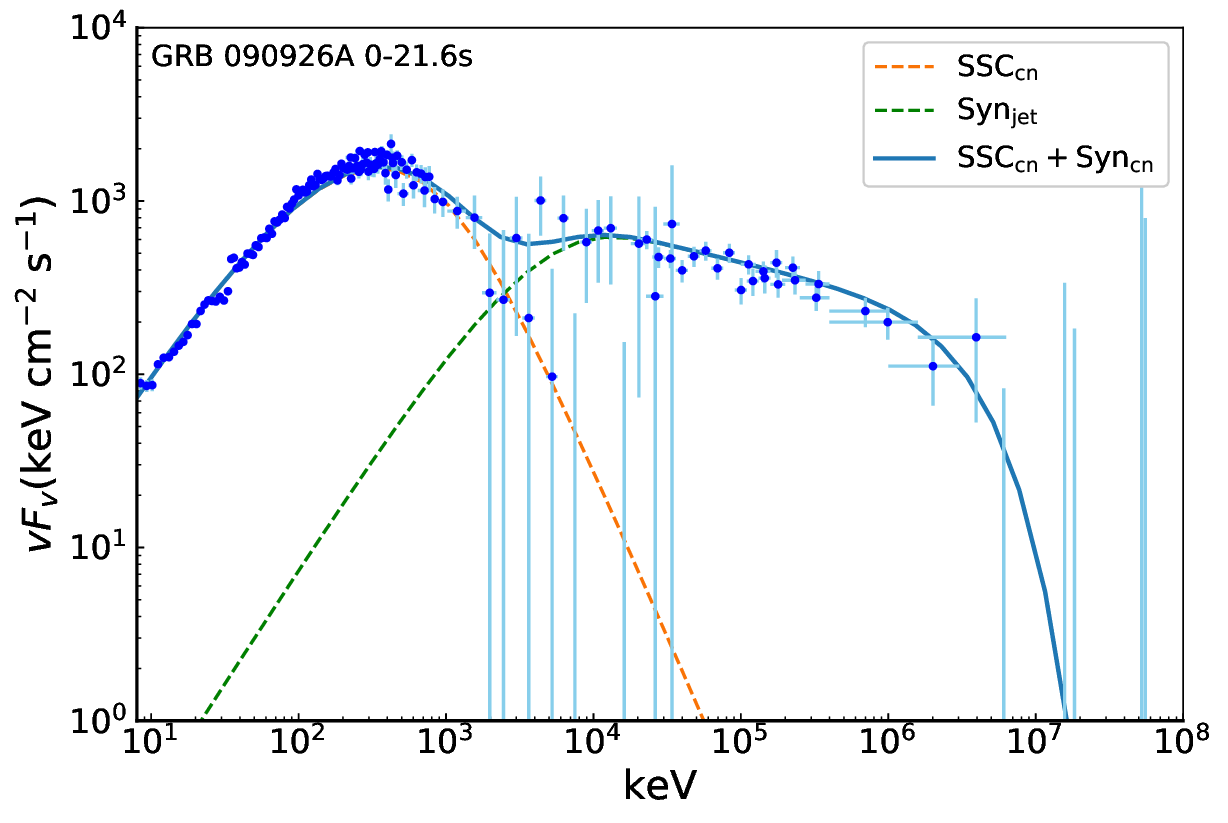}
        \includegraphics[width=0.55\linewidth]{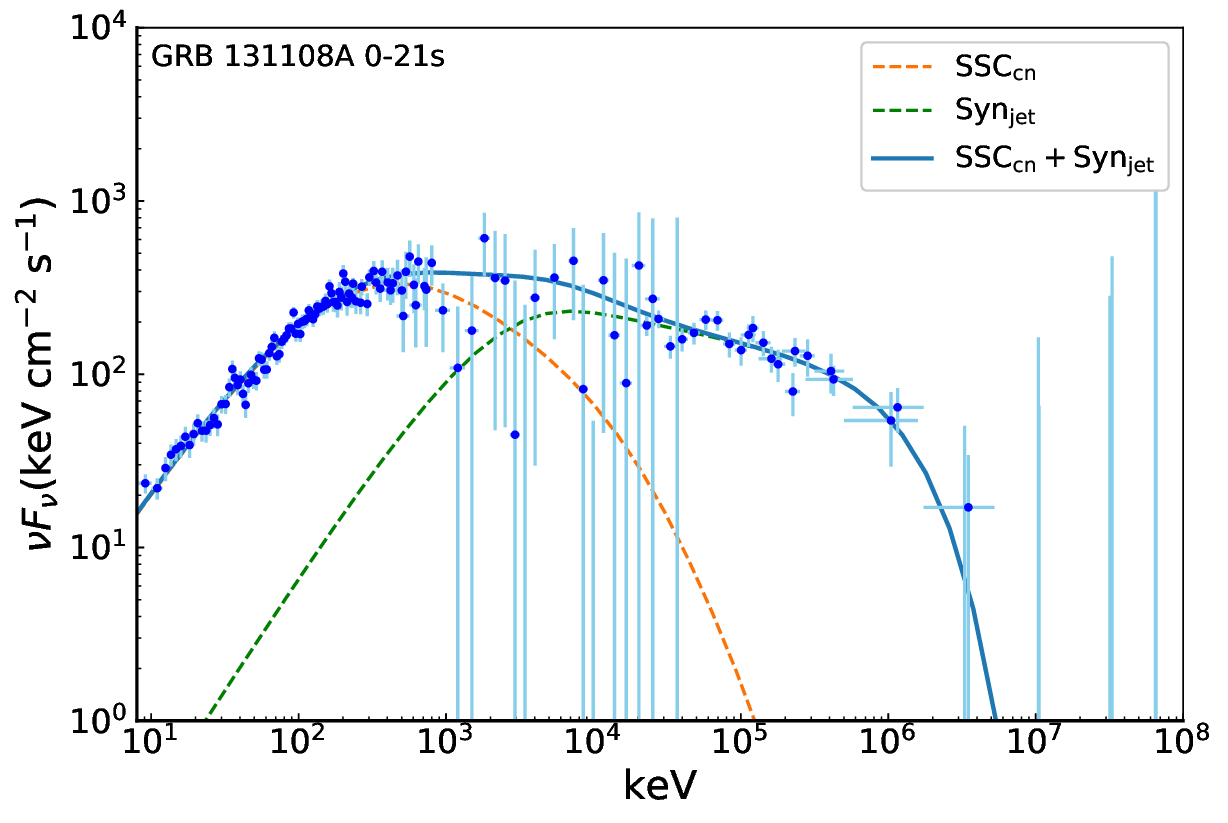}
        \includegraphics[width=0.55\linewidth]{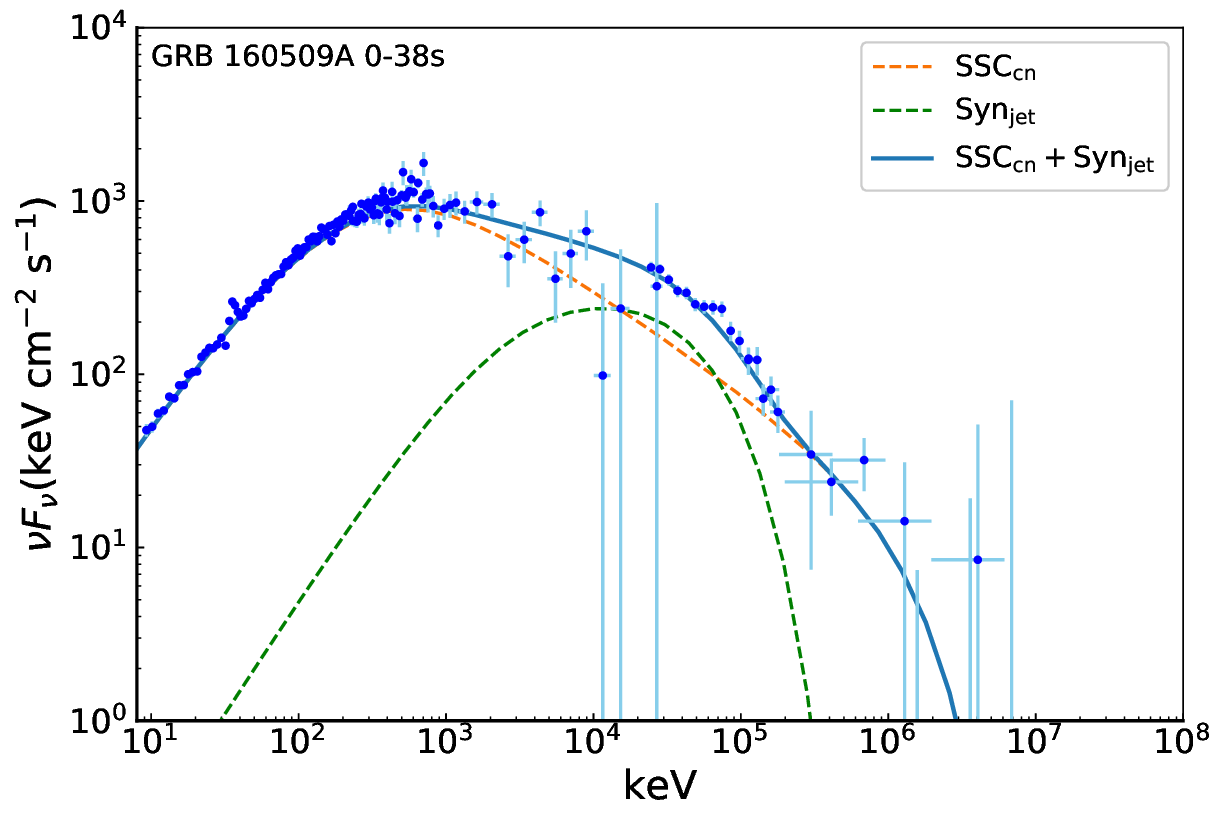 }
    \caption{Time-integrated spectra of GRBs 090926A, 131108A, and 160509A, along with theoretical fits by our model (solid lines). The emission components of the MJC and the jet core regions are marked with dashed lines.}
    \label{fig:all case}
\end{figure}

The spectral characteristics predicted by our model can potentially accommodate the diversity of the prompt gamma-ray spectra observed with telescopes across different energy bands. In this section, we apply our model to three GRBs (GRBs 090926A, 131108A, and 160509A) whose spectra exhibit a Band-cut or saddle shape. We download the GBM and LAT data of the three GRBs from the public science support center on the official $Fermi$ Web site\footnote{\url{http://fermi.gsfc.nasa.gov/ssc/data/}}. The GBM comprises 12 sodium iodide (NaI) detectors covering an energy range from 8 keV to 1 MeV, and two bismuth germanate (BGO) scintillation detectors sensitive to higher energies between 150 keV and 40 MeV. We select the brightest NaI and BGO detectors. The LAT is a pair conversion telescope with energy coverage ranging from below 20 MeV to over 300 GeV. Data reduction is performed using the Fermitools-v2.2.0 package and the P8\_TRANSIENT020E response function. We extract the time-integrated spectra of these GRBs with the GBM and LAT data and fit the data with our model. The observed spectra and our fits are shown in Figure~\ref{fig:all case}. The derived model parameters are listed in Table~\ref{tab:parameters}. We describe the results below. 

\begin{figure}[htbp!]
    \centering
        \includegraphics[width=0.55\linewidth]{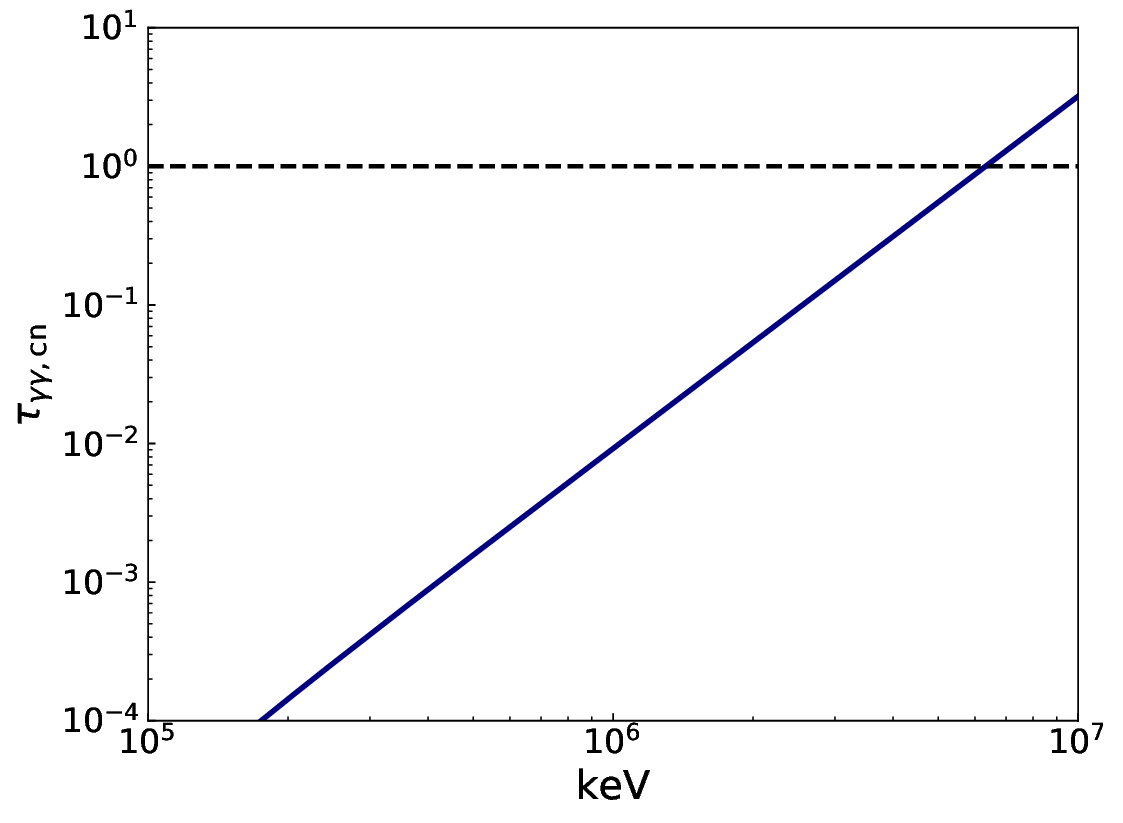} 
    \caption{The optical depth for $\gamma \gamma $ annihilation of gamma-ray photons in the MJC region of GRB 160509A as a function of photon energy. }
    \label{fig:optical}
\end{figure}

\begin{itemize}
    \item GRB 090926A: It is a bright, long burst at redshift of $z=2.1062$ \citep{2009GCN..9942....1M}. Its $\mathrm{T}_{90}$ duration measured with GBM is approximately 21 s \citep{2009GCN..9972....1B}. The derived time-integrated spectrum is accumulated from $T_0$ to $T_0+21.6$ s. It clearly shows a saddle shape. The initial Lorentz factor of its jet core is $\Gamma_{\rm jet} \sim 600$ \citep{2011ApJ...729..114A}. The observed spectrum can be well represented by our model. The bright peak at several hundred of keV is attributed to the $\rm SSC_{cn}$-component and the broad hump in 10 MeV$-$10 GeV is dominated by the $\rm Syn_{jet}$-component. Compared with the jet core, the cocoon is sub-relativistic ($\beta_{\rm cn, 0}=0.83$ vs. $\Gamma_{\rm jet}=611$) and low magnetization ($B_{\rm cn}=54$ vs. $B_{\rm jet}= 1 \times 10^6$ G).  
    
    \item GRB 131108A: It is also a bright burst at $z\sim 2.40$ \citep{2019ApJ...886L..33A}. Its initial Lorentz factor of the jet core is set as $\Gamma_{\rm jet} \sim 500$ \citep{2018A&A...609A.112G}. Its SED illustrates a typical Band-Cut function. Our model can well fit the spectrum. Similar to that of GRB 090926A, the SED of GRB 131108A in the keV$-$MeV band is contributed by the $\rm SSC_{cn}$-component, and the broad bump at the MeV$-$GeV band is dominated by the $\rm Syn_{jet}$-component. The derived model parameters are also similar to those of GRB 090926A. 

    \item GRB 160509A: It is a bright GRB at $z \approx 1.17$ \citep{2016GCN.19419....1T}. The prompt emission light curve can be segmented into three distinct phases: a soft ``precursor" peak ($T_0-5.0$ s $\sim T_0+5.0$ s), a bright main episode ($T_0+5.5$ s $\sim T_0+37$ s), and a subsequent weak emission episode ($T_0+300$ s $\sim T_0+400$ s) \citep{2018ApJ...864..163V}. We focus on the radiation characteristics during the primary emission episode that spans from $T_0$ to $T_0+38$ s. The initial Lorentz factor of the jet is set as $\Gamma _{\rm jet} \sim 300$ \citep{2016ApJ...833...88L}. Its SED closely resembles the SED of GRB 131108A at $E<30$ MeV, but has a power-law spectrum in the range from 30$-$300 MeV range. This power-law decaying segment even extends beyond 1 GeV. We fit the SED with our model and find that the sum of the $\rm SSC_{cn}$- and the $\rm Syn_{jet}$- components can well represent the observed SED. The $\rm SSC_{cn}$-component almost dominates the observed in the keV$-$MeV$-$GeV band. The $\rm Syn_{jet}$-component peaks around 10 MeV, with a peak flux being comparable to the $\rm SSC_{cn}$-component. The $\rm SSC_{cn}$-component peaks at $\sim 300$ keV and decays as a power-law up to 3 GeV. The emission above $\sim 0.1$ GeV is attributed to the high-energy tail of the $\rm SSC_{cn}$-component. The $B_{\rm cn}$ of GRB 160509A is $5\sim 8$ times larger than that of GRB 090926A and GRB 131108A. The gamma-ray emission of GRB 160509A in the energy band beyond 100 MeV is attributed to the emission from the $\rm SSC$ process in the MJC region. We examine the optical depth of the gamma-rays for $\gamma \gamma $ annihilation ($\tau_{\gamma \gamma, \rm cn }$) in the MJC region of GRB 160509A. The result is shown in Figure~\ref{fig:optical}. It is shown that $\tau_{\gamma \gamma, \rm cn } < 1$ at $ < 6$ GeV, but it is larger than 1 beyond this energy range. The detection of gamma-rays at $E \sim 3$ GeV in GRB 160509A agrees with the transparency condition.  
\end{itemize}

\begin{table}[htbp!]
    \centering
    \caption{{Model parameters derived from our fits to the observed SEDs.}}
    \begin{tabular}{cccc|ccccccc}
        \hline\hline
        Name & $\beta _{\mathrm{cn},0}$ & $B_{\mathrm{cn}}$ & $\gamma _{e, \mathrm{inject}}$($\Gamma_{\rm jet}$) & $B_{\rm jet}$ & $p_{\rm jet}$ & $\gamma_{m, \rm jet}$ & $\gamma_{\rm b,jet}$ & $\gamma _{\rm M, jet}$ \\
        \hline
        GRB 090926A & $0.830$ & $54 $ & $6.11 \times 10^{2}$ & $1 \times 10^{6}$ & $2.4$ & $6.5 \times 10^{3}$ & $9.8 \times 10^{3}$ & $2 \times 10^{5}$ \\
        \hline
        GRB 131108A & $0.895$ & $80 $ & $5.02 \times 10^{2}$ & $1 \times 10^{6}$ &$2.35$ & $5 \times 10^{3}$ & $7 \times 10^{3}$ & $1.3 \times 10^{5}$ \\
        \hline
        GRB 160509A & $0.906$ & $450 $ & $3.31 \times 10^{2}$ & $3 \times 10^{6}$ & $2.1$ & $5 \times 10^{3}$ & $1.3 \times 10^{4}$ & $2.5 \times 10^{4}$\\
        \hline
    \end{tabular}
    \label{tab:parameters}
\end{table}


\section{Summary and Discussion} \label{sec:Summary}
Assuming that the GRB jet is structured, we propose that the observed GRB spectrum in the keV$-$GeV band is attributed to emission from the electrons accelerated via internal shocks in the relativistic jet core and the electrons accelerated through the shear acceleration mechanism in the sub-relativistic MJC region. Taking $\gamma_{\rm e, inject}=300$, $B_{\rm cn}=100$ G, we show that the $\rm SSC$ process governs the cooling of electrons in the MJC region. The $\rm SSC_{cn}$ emission flux below $10^{19}$ Hz is almost independent of $\beta_{\rm cn,0}$ but it is sensitive to $\beta_{\rm cn,0}$ at $\nu>10^{19}$ Hz. Combing both the emission from the jet core and the MJC region, the overall SED in the keV$-$MeV$-$GeV band shape as a Band function or a Band-cut function with an X-ray excess if $\beta_{\rm cn,0}\lesssim 0.9$, $B_{\rm cn} \sim 10^{2}$ G, $\theta_{\rm cn}=0.7$ rad, $\Gamma_{\rm jet} \sim 300$, $B_{\rm jet}\sim 10^{6}$ G, and $\theta_{\rm jet}=0.07$ rad. We apply our model to explain the prompt gamma-ray spectra of bright GRBs 090926A, 131108A, and 160509A whose spectra distinctly show two components or a Band-cut function shape. We show that these spectra can be effectively explained with our model. 

\begin{figure}[htbp!]
    \centering
    \includegraphics[width=0.32\textwidth]{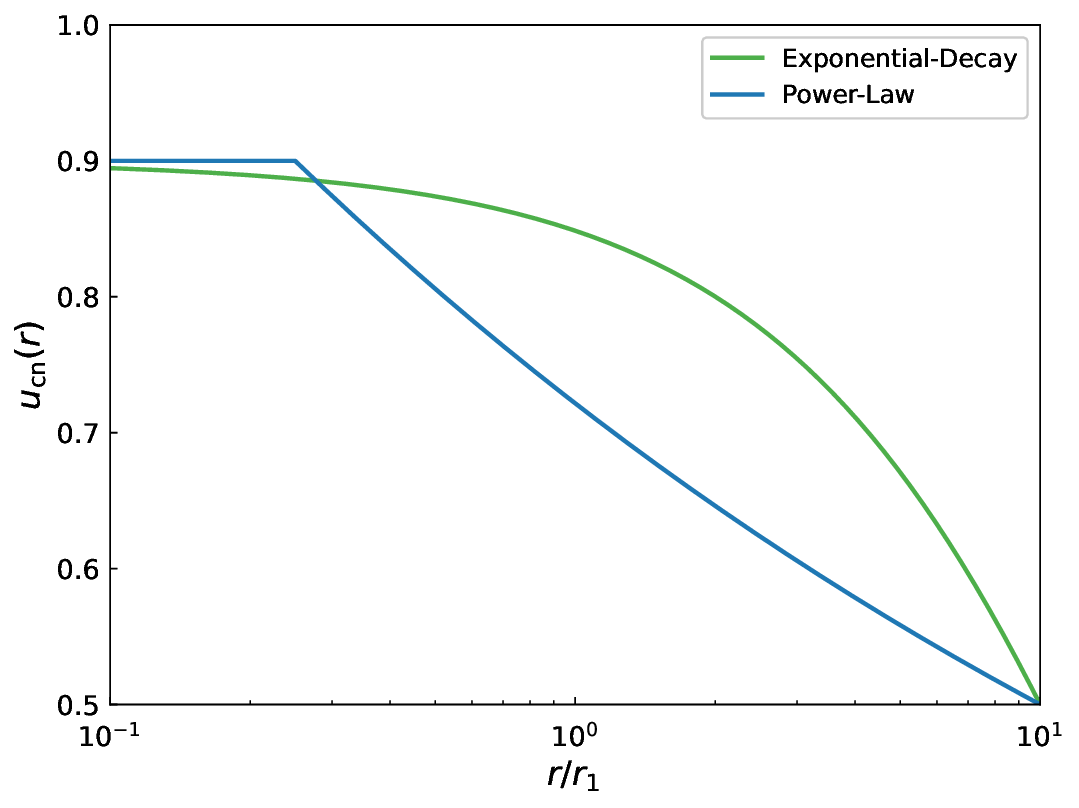}
    \includegraphics[width=0.32\textwidth]{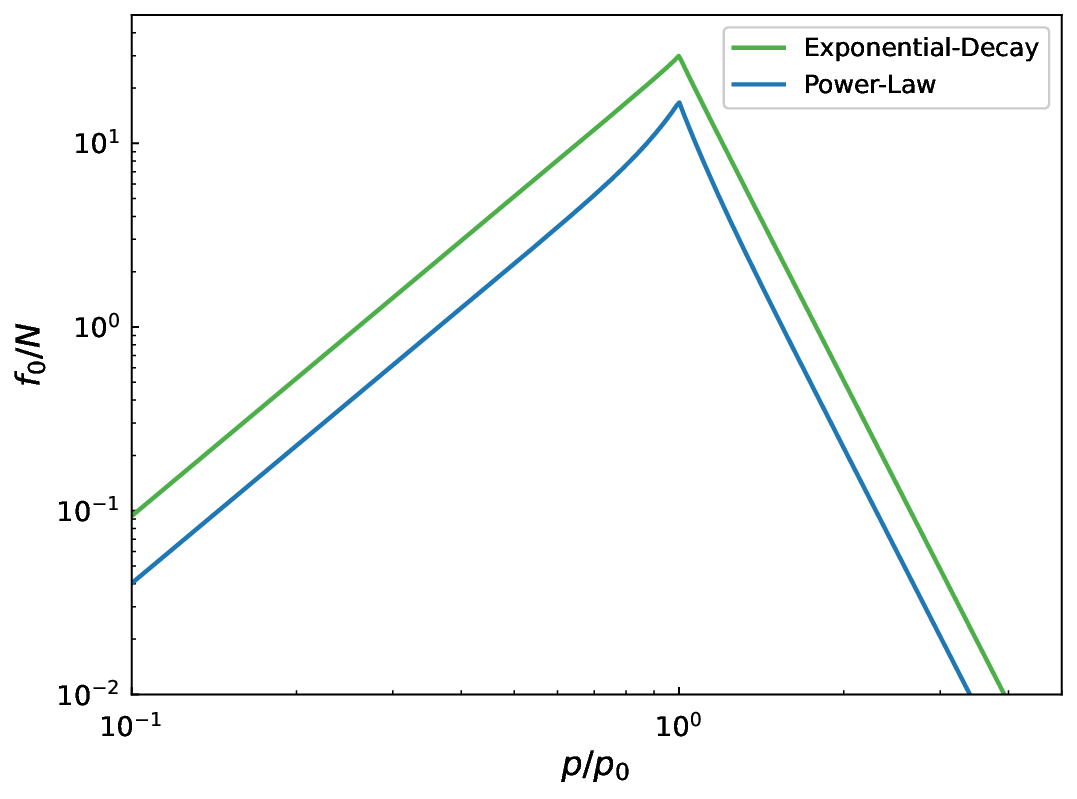}
    \includegraphics[width=0.32\textwidth]{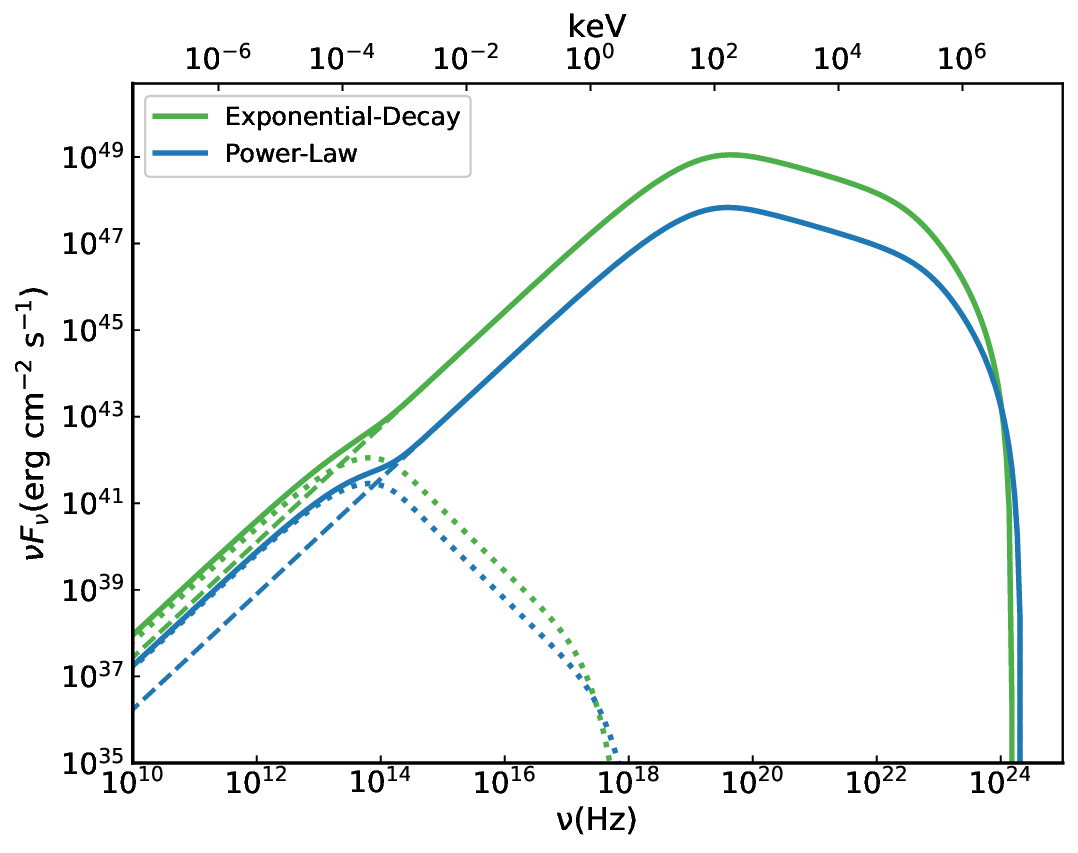}
    \caption{{\rm Left panel--} Velocity profiles of the MJC region as exponential-decay and power-law functions of radius, with the initial velocity of $\beta_{\rm cn,0}=0.9$. {\rm Right panel--} Distributions of shear-accelerated electrons corresponding to the velocity profiles in the left panel.}
    \label{fig:compare}
\end{figure}

In this paper, we employ the exponential-decay function to represent the radial velocity profile. Nevertheless, the actual velocity profile may be described as other characteristic functions. We additionally consider a scenario where the velocity profile follows a power-law function and compare the corresponding electron distribution with the exponential-decay case. Figure~\ref{fig:compare} presents the comparison results using the same parameter set outlined above, with the initial velocity of $\beta_{\rm cn,0}=0.9$. The results indicate that the structural morphology of shear-accelerated electron distributions persists across different profiles, as described in Eq.~\ref{eq:10} and Eq.~\ref{eq:11}. Variations in the derivative of the velocity profile function affect the efficiency of the shear acceleration process, resulting in differences in the electron distribution. Nonetheless, the primary conclusions of this article remain unaffected.

An X-ray excess over the Band function in the several keV bands has been detected in some GRBs observed with CGRO/BATSE. \cite{1996ApJ...473..310P} analyzed time-averaged spectra from 86 bright GRBs observed during the first five years of BATSE and found that 12 bursts exhibit an excess of low-energy emission in the 5$-$20 keV range, with a significance exceeding 5$\sigma$. A similar signature is also observed with the $\rm Ginga$ observation at energies as low as 2 keV \citep{1998ApJ...500..873S}.
It is uncertain whether the X-ray excess is the tip of an ice-burger of the photosphere emission of the ``hot" fireball. Inspecting the SEDs shown in Figure~\ref{fig:spectrum}, one can observe that the sum of the $\rm SSC_{cn}$-component and the $\rm Syn_{jet}$-component produce a bump-like feature, which can mimic as an X-ray excess over the fitting curve with the Band function. 

\begin{acknowledgements}
This work is supported by the National Natural Science Foundation of China (Grant Nos. 12203015, 12133003). This work is also supported by the Guangxi Talent Program (“Highland of Innovation Talents”), and the startup financial support program of Guizhou Normal University (grant No. GZNUD[2023]).
\end{acknowledgements}

\bibliography{sample631}{}
\bibliographystyle{aasjournal}

\end{document}